%% file: IteratedCPMs.tex
\documentclass[submission,copyright,creativecommons,sharealike,noncommercial]{eptcs}
\providecommand{\event}{QPL 2020}
\pdfoutput=1

\usepackage{graphicx}
\usepackage{subcaption}
\usepackage{dcolumn}
\usepackage{bm}
\usepackage{hyperref}
\usepackage{tikz-cd}
\usepackage{tikz}
\usepackage{bbm}
\usepackage{physics}
\usepackage{dsfont}
\usepackage{float}
\usepackage{amsthm}
\usepackage{circuitikz}
\usepackage{tikzit}
\usepackage{url}
\usepackage{relsize}
\usepackage{amsmath}
\usepackage{amsfonts}
\usepackage{amssymb}

\input{tikzstyles}

\input{styles.tikzstyles}
\usetikzlibrary{calc, positioning, shapes.geometric}
\usetikzlibrary{
	arrows,
	shapes,
	decorations,
	intersections,
	backgrounds,
	positioning,
	circuits.ee.IEC
	}

\newcommand{\tikzfigscale}[2]{\scalebox{#1}{\input{#2.tikz}}}

\graphicspath{{./figs/}}

\newcommand{\cat}{\mathbf}

\newcommand{\morph}[1]{\xrightarrow{#1}}

\newcommand{\fdhilb}{\textbf{FdHilb}}

\newcommand{\cpm}[1]{\textbf{CPM}(#1)}

\newcommand{\dbl}[1]{\textrm{dbl}(#1)}
\newcommand{\fld}[1]{\textrm{fld}(#1)}

\newcommand{\dec}[1]{\textrm{dec}_{#1}}
\newcommand{\hypdec}[1]{\textrm{hypdec}_{#1}}

\newcommand{\dscfa}{\dag\textrm{-SCFA}}

\newcommand{\UHfB}[1]{%
\mathbin{\tikzfigscale{#1}{figs/UHfB}}%
}

\newcommand{\littlestate}[1]{%
\mathbin{\text{\begin{tikzpicture}
		\node [style=state, scale=0.7] (0) at (0, 0) {$#1$};
		\node [style=none] (1) at (0, 0.3) {};
		\draw (0) to (1);
\end{tikzpicture}}}%
}

\newcommand{\littleantipode}{%
\mathbin{\text{\begin{tikzpicture}
		\node [style=none] (0) at (0, -0.3) {};
		\node [style=antipode, scale=0.5] (1) at (0, 0) {};
		\node [style=none] (2) at (0, 0.3) {};
		\draw (0.center) to (1);
		\draw (1) to (2.center);
\end{tikzpicture}}}%
}

\theoremstyle{definition}
\newtheorem{defn}{Definition}
\theoremstyle{plain}

\theoremstyle{plain}

\theoremstyle{plain}
\newtheorem{prop}{Proposition}

\begin{document}

\title{Hyper-decoherence in Density Hypercubes}
\author{
    James Hefford \institute{University of Oxford} \email{james.hefford@cs.ox.ac.uk}
    \and
    Stefano Gogioso \institute{University of Oxford} \email{stefano.gogioso@cs.ox.ac.uk}
}

\def\titlerunning{Hyper-decoherence in Density Hypercubes}
\def\authorrunning{J. Hefford and S. Gogioso}

\maketitle


\begin{abstract}
	We study hyper-decoherence in three operational theories from the literature, all examples of the recently introduced higher-order CPM construction.
	Amongst these, we show the theory of density hypercubes to be the richest in terms of post-quantum phenomena.
	Specifically, we demonstrate the existence of a probabilistic hyper-decoherence of density hypercubes to quantum systems and calculate the associated hyper-phase group.
	This makes density hypercubes of significant foundational interest, as an example of a theory which side-steps a recent no-go result in an original and unforeseen way, while at the same time displaying fully fledged operational semantics.
\end{abstract}

\section{Introduction}

Interference is a fundamental feature of quantum theory, responsible for advantage in a number of computational tasks.
However, quantum interference is known to be limited to the \emph{second order}: the pattern produced by three or more slits can always be classically explained as a combination of the patterns arising from the individual slits and the pairs of slits.
In previous literature, higher-order interference has been considered as a hallmark of post-quantum theories.
In particular, Sorkin proposed that theories be hierarchically organised based on the maximum order of interference they exhibit \cite{sorkin_measure,Sorkin2}.
More recent work has explored the operational implications of higher-order interference in such theories \cite{lee_extensions}.

Another feature considered characteristic of post-quantum theories is the necessity for quantum theory to arise as an effective sub-theory, by some ``higher-order'' analogue of decoherence---the mechanism by which classical theory arises as an effective sub-theory of quantum theory.
The possibility of such a \emph{hyper-decoherence} process happening within operational theories of nature has been explored in \cite{lee_nogo}, with a no-go result limiting its ultimate scope of applicability.

The first fully fledged probabilistic theory displaying higher-order interference and hyper-decoherence was recently introduced in \cite{gogioso_hyper} under the name of \emph{density hypercubes}, based on a higher-order extension of the construction used to obtain mixed-state quantum theory from pure state quantum theory \cite{gogioso_cpm}.
While the original \cite{gogioso_hyper} introduced suitable hyper-decoherence maps, it failed to show that such maps had a well-defined operational interpretation: it was known that they would happen probabilistically as part of some larger process, but it was not known what that process could look like.

In this work we patch the short-comings of \cite{gogioso_hyper} and put density hypercubes on solid footing as a probabilistic theory, showing that hyper-decoherence truly has a bona fide operational interpretation as a probabilistic component of a larger deterministic process.
This will allow further operational exploration of post-quantum effects in density hypercubes to be carried out with the necessary confidence in its theoretical foundations.

This work also explores two closely related theories from the literature, those of \emph{double dilation} and \emph{double mixing}, developed to describe quantum-like aspects of ambiguity in natural language processing \cite{coecke_double, piedeleu}.
It was originally believed \cite{gogioso_hyper} that density hypercubes and double dilation coincided: we show that not to be the case. We further show that double dilation and double mixing do not feature the hyper-decoherence maps of density hypercubes, nor the same associated phase group.

Here follows a brief summary of known results about double dilation, double mixing and density hypercubes, together with open questions which this work addresses.

\begin{itemize}
    \item It was known that double mixing is a sub-theory of double dilation \cite{coecke_dilation}. A straightforward adaptation of the same argument also shows that double mixing is a sub-theory of density hypercubes.
    \item It was originally believed that density hypercubes and double dilation were the same theory \cite{gogioso_hyper}. This work will show that not to be the case.
    \item It was known that density hypercubes show higher-order interference, that their Karoubi envelope contains both quantum theory and classical theory, and that there exist hyper-decoherence maps from hypercubes to quantum theory \cite{gogioso_hyper}.
    It was not known whether the hyper-decoherence maps were sub-normalised, i.e. whether they were a probabilistic outcome of some larger deterministic process of density hypercubes. The main achievement of this work will be to show that this is indeed the case.
    \item The structure of the hyper-phase group---the invertible maps which are quotiented away by hyper-decoherence---for density hypercubes was not known. This work will show that there exist non-trivial elements of this group.
    \item It was not known whether double dilation and double mixing possess hyper-decoherence maps, or what the associated hyper-phase group would be. This work will show that the same choice of maps from density hypercubes would not work.
    \item Although the above is not a complete answer, this work will provide evidence against the existence of such hyper-decoherence maps by showing that severe restrictions apply to the maps that the associated hyper-phase group can contain.
\end{itemize}

\section{Iterated CPM Constructions}

The CPM construction \cite{selinger} is the categorical way of capturing the passage from pure quantum mechanics (in terms of vectors in a Hilbert space) to mixed state quantum mechanics (where states are associated with operators on the Hilbert space).
At its core, the construction can be understood in two steps.
Firstly, we ``double'' the morphisms of the original (dagger compact) category:

\begin{equation}\label{fold}
	\tikzfigscale{1}{figs/fold}
\end{equation}

\noindent
Secondly, we allow for discarding of environment systems:

\begin{equation}\label{cpm}
    \tikzfigscale{1}{figs/cpmaps}
    \hspace{10mm} \raisebox{0mm}{=} \hspace{10mm}
	\tikzfigscale{1}{figs/environment}
\end{equation}

\noindent
The discarding is defined in terms of the \emph{cap} arising from the duality of $E$ and $E^*$ and can be thought of as a categorical generalisation of the partial trace.
If the underlying category is $\fdhilb$ (the dagger compact category of finite dimensional Hilbert spaces and complex linear maps), then the maps in the shape of \eqref{cpm} are exactly the completely positive maps $\dbl{H} \rightarrow \dbl{K}$, where we have defined the \emph{doubling functor} $\dbl{H} := H \otimes H^*$ and $\dbl{f} := f \otimes f^*$.
In particular, the states $I \rightarrow \dbl{H}$ are Choi-Jamio\l{}kowski isomorphic to positive operators $H \rightarrow H$.
The dagger compact category with objects in the form $\dbl{H}$ for some finite-dimensional Hilbert space $H$ and morphisms in the shape of \eqref{cpm} is called $\cpm{\fdhilb}$.

A question now arises: Can the CPM construction be iterated?
If so: What phenomena do the resulting theories exhibit?
The second author recently described \cite{gogioso_cpm} how the CPM construction can be iterated and generalised by introducing a ``folding'' functor---generalising the doubling---and a ``multi-environment structure''---generalising the discarding.
In such a generalised CPM construction, one starts with a group action $\Phi:G\morph{}\textrm{Aut}(\cat{C})$ by monoidal functors of some finite abelian group $G$ on a symmetric monoidal category $\cat{C}$.
The folding functor is defined on objects by $\fld{A} := \bigotimes_{\gamma\in G} \Phi(\gamma)[A]$ and on morphisms by $\fld{f} := \bigotimes_{\gamma\in G} \Phi(\gamma)[f]$.
If we take the group $G = \mathbb{Z}_2$ and consider the action defined by $\Phi(1) := \mapsto \operatorname{conj}_\cat{C}$ (where $\mathcal{C}$ is a dagger compact category and $\operatorname{conj}_\cat{C}$ is the dualising functor) we recover the original doubling: on objects we get $ \fld{A} = A\otimes A^*$ and on morphisms we get $\fld{f} = f\otimes f^* = \dbl{f}$ on morphisms.
The original CPM construction can be recovered with the multi-environment structure $\Xi_A := \{\epsilon_A\}$, where $\epsilon_A: A\otimes A^* \rightarrow I$ is the cap.

\section{Double dilation, double mixing and density hypercubes}

In this work we will focus on three cases where the group $G$ used for folding is $\mathbb{Z}_2\cross\mathbb{Z}_2$, with action defined by $\Phi(1,0) = \Phi(0,1) := \mapsto \operatorname{conj}_\cat{C}$; that is, we iterate the original CPM construction twice.
Explicitly, the folding functor sends objects $H$ to objects $H \otimes H^* \otimes H \otimes H^*$, and similarly for morphisms.
Different choices of multi-environment structures then capture different ways of discarding: the three examples we consider all contain the doubling of the discarding map from $\cpm{\fdhilb}$ plus one additional effect.

The first choice appearing in the literature is that of \textit{double dilation} \cite{coecke_dilation}, also known as \textit{dual density operators} \cite{ashoush}.
The multi-environment structure contains the two possible ways in which caps $\epsilon_A: A^*\otimes A \rightarrow I$ can be applied to $\dbl{\dbl{E}} = E \otimes E^* \otimes E \otimes E^*$: (i) the doubling $\dbl{\epsilon_{E}}$ of the discarding maps from $\cpm{\fdhilb}$ and (ii) the cap $\epsilon_{\dbl{E}}$:

\begin{equation}\label{doubledil}
    \tikzfigscale{1}{figs/ddmaps}
    \hspace{10mm} \raisebox{0mm}{=} \hspace{10mm}
	\tikzfigscale{1}{figs/doubledilation}
\end{equation}

\noindent
Throughout this work, we will use thick wires and borders to indicate diagrams living in the category $\cpm{\fdhilb}$ and thin wires and boxes for diagrams living in the category $\fdhilb$.

The second choice appearing in the literature is that of \textit{double mixing} \cite{coecke_dilation}.
The multi-environment structure contains the doubling $\dbl{\epsilon_{E}}$ of the discarding maps from $\cpm{\fdhilb}$ together with a four-legged spider connecting all four environment systems:

\begin{equation}\label{doublemix}
    \tikzfigscale{1}{figs/dmmaps}
    \hspace{10mm} \raisebox{0mm}{=} \hspace{10mm}
	\tikzfigscale{1}{figs/doublemixing}
\end{equation}

\noindent
In the above, \raisebox{-1pt}{\scalebox{1.5}{$\circ$}} is a special commutative $\dagger$-Frobenius algebra ($\dscfa$) \cite{coecke_bases}.
It can be shown \cite{coecke_dilation} that double mixing is a sub-theory of double dilation, i.e. that all maps in the form \eqref{doublemix} can also be put in the form \eqref{doubledil}.
Both double dilation and double mixing have found application to the modelling of ambiguity in natural language processing \cite{coecke_double, piedeleu}.

The third choice appearing in the literature is that of \textit{density hypercubes} \cite{gogioso_hyper}.
The multi-environment structure contains the doubling $\dbl{\epsilon_{E}}$ of the discarding maps from $\cpm{\fdhilb}$ together with two-legged spiders resembling the cap $\epsilon_{\dbl{E}}$ from double dilation:

\begin{equation}\label{hypercube}
    \tikzfigscale{1}{figs/dhmaps}
    \hspace{10mm} \raisebox{0mm}{=} \hspace{10mm}
	\tikzfigscale{1}{figs/densityhypercubes}
\end{equation}

\noindent
Despite the apparent similarity, density hypercubes are significantly different from double dilation: the two-legged spiders used in the RHS of \eqref{hypercube} are effects on $E \otimes E$ and $E^* \otimes E^*$, while the two caps forming $\epsilon_{\dbl{E}}$ in \eqref{doubledil} are both effects on $E \otimes E^*$.
Because of this, the objects of density hypercubes are most naturally written in the form $\dbl{H} \otimes \dbl{H}$ (instead of $\dbl{H^*} \otimes \dbl{H}$) and the folded morphisms in density hypercubes are most naturally written in the form $\dbl{f} \otimes \dbl{f}$ (instead of $\dbl{f^*} \otimes \dbl{f}$).
This will turn out to make a very significant difference.

On the LHS of \eqref{hypercube}, the wires have been re-arranged to achieve a more pleasant visual effect. In more direct analogy with double dilation and double mixing they would have taken the following form:

\begin{equation}\label{hypercubeHorrid}
    \tikzfigscale{1}{figs/dhmaps}
    \hspace{7.5mm} \raisebox{3mm}{:=} \hspace{7.5mm}
    \tikzfigscale{1}{figs/dhmapsHorrid}
\end{equation}

\noindent
We refer to the two legged spider appearing in the middle as a \emph{bridge}.
For convenience, we will allow for bridges in any choice of $\dscfa$ \raisebox{-1pt}{\scalebox{1.5}{$\circ$}}: this generalisation yields the same class of morphisms, but with added flexibility when drawing diagrams.

In the generalised CPM construction, a single family of effects has to be chosen within the multi-environment structure to endow the theory with a notion of causality and normalisation \cite{chiribella_purification,coecke2013causal,coecke2016terminality}.
We make the same choice for all three theories:

\begin{equation}
	\tikzfigscale{1}{figs/discarding}
\end{equation}

\noindent
We refer to the effects above as the \emph{discarding maps} for the theories. The \emph{normalised} morphisms in the three theories are those which respect the choice of discarding maps made above, in the following sense:

\begin{equation}
	\tikzfigscale{0.6}{figs/causalitydd}
    \hspace{10mm}
    \tikzfigscale{0.6}{figs/causalitydm}
    \hspace{10mm}
    \tikzfigscale{0.6}{figs/causalitydh}
\end{equation}

\noindent
The three theories are also \emph{probabilistic}: they have the non-negative real numbers $\mathbb{R}^+$ as scalars, with morphisms that can be rescaled and added together.
In particular, normalised morphisms form a convex set and are interpreted as processes that can be made to happen ``with certainty'', or ``deterministically''.
More generally, we say that a morphism $f: H \rightarrow K$ is \emph{sub-normalised} if there is some $g: H \rightarrow K$ such that $f + g$ is normalised: sub-normalised morphisms are interpreted as processes that happen ``probabilistically''---with probability dependent on the specific state that they are applied to---as cases of a larger deterministic process.
There is a unique normalised scalar, the number $1$, and sub-normalised scalars coincide with probabilities $[0, 1]$.

\section{Hyper-decoherence in probabilistic theories}

In quantum theory, the process of \emph{decoherence} leads to the emergence of classical theory and is given by the following map \cite{coecke_environment}:

\begin{equation}
    \tikzfigscale{1}{figs/quantdec}
\end{equation}

\noindent
Such a map acts to zero out the non-diagonal entries of a density matrix in the basis associated with the $\dscfa$ \raisebox{-1pt}{\scalebox{1.5}{$\circ$}} \cite{coecke_bases}: a quantum state is sent to a classical probability distribution.
Decoherence maps are normalised---i.e. they correspond to processes the can be made to happen with certainty---and idempotent: once a quantum system is decohered to a classical system, further applications of the decoherence map produce no additional effect.

\emph{Hyper-decoherence} is analogous to decoherence, but one level up: it leads to the emergence of quantum theory from some post-quantum (or \emph{hyper}-quantum) theory, by suppression of the hyper-quantum part.
The existence of hyper-decoherence maps has been considered in the literature as a possible mechanism for our lack of observation of hyper-quantum effects \cite{quarticquantum, density_cubes, lee_extensions}: perhaps we simply cannot perform experiments accurately  enough to see such effects, or perhaps hyper-decoherence happens on time scales shorter than those currently accessible to experimentalists.
If quantum theory is to be deemed an effective---as opposed to fundamental---theory of nature, hyper-decoherence is one possible mechanism to explain why we have yet to observe post-quantum phenomena.

In the literature, hyper-decoherence maps have been defined analogously to decoherence maps in quantum theory: idempotent and normalised maps taking hyper-quantum states to quantum states.
A known no-go result \cite{lee_nogo} states that such hyper-decoherence maps cannot exist in operational probabilistic theories with purification if some additional assumptions are imposed---namely that pure states in quantum theory be pure in the larger post-quantum theory and that the maximally mixed state of quantum theory be maximally mixed in the larger post-quantum theory.
While at the time those assumptions were deemed operationally sensible, the discovery of hyper-decoherence maps in density hypercubes requires a broadening of the definition.

\begin{defn}
    Given any probabilistic theory \cite{gogioso_cpt}, the \emph{Karoubi envelope} is the probabilistic theory having objects in the form $(H, e)$, where $H$ is some object in the original theory and $e: H \rightarrow H$ is an idempotent.
    Processes $F: (H, e) \rightarrow (H', e')$ in the Karoubi envelope are exactly the processes $F: H \rightarrow H'$ in the original theory which are invariant under the idempotents, i.e. such that $F = e' \circ F \circ e$. \cite{selinger_idempotents, coecke_classicality, gogioso_cpt}
\end{defn}

\noindent
From an operational perspective, objects $(H, e)$ of the Karoubi envelope capture a situation in which it can be safely assumed that an idempotent process $e$ has taken place between any two operations, e.g. because it happens on time-scales much smaller than those operationally accessible.
This is, for example, the way in which classical systems arise from quantum systems by decoherence.

\begin{defn}
    In a probabilistic theory, a \emph{hyper-decoherence map} is a sub-normalised idempotent process $\operatorname{hypdec}: H \rightarrow H$ such that the object $(H, \operatorname{hypdec})$ in the Karoubi envelope is a quantum system.
    A probabilistic theory is \emph{post-quantum} if there is a full sub-category of its Karoubi envelope spanned by hyper-decoherence maps which is equivalent to quantum theory, i.e. to $\cpm{\fdhilb}$.
\end{defn}

From an operational perspective, a post-quantum theory is one such where quantum theory arises as an effective theory by means of hyper-decoherence happening \emph{probabilistically} at time-scales much smaller than those operationally accessible to quantum experiments.
Idempotence of hyper-decoherence maps ensures that once a systems has collapsed to quantum it remains quantum.
Idempotence also ensures that the probabilistic nature of hyper-decoherence manifests exactly once: conditional hyper-quantum collapse having happened at least once, hyper-decoherence is deterministic and does nothing to the quantum system.
If observers are for some reason limited to the quantum part of the theory, hyper-decoherence would happen transparently to them: this is not too far removed from what is speculated to happen in string theory and brane cosmology, where the observable world is restricted to a brane within a larger bulk.

\section{Hyper-decoherence in density hypercubes}

It has been shown \cite{gogioso_hyper} that the theory of density hypercubes has both decoherence maps (collapse to classical theory) and hyper-decoherence maps (collapse to quantum theory), taking the following form:

\begin{equation}
    \tikzfigscale{1}{figs/dechypdec}
\end{equation}

\noindent
Both maps are idempotent, but unfortunately they are not normalised, meaning that they are not---on their own---bona fide physical processes:

\begin{equation}\label{hypdecsubcaus}
    \tikzfigscale{1}{figs/hypdecsubcaus}
\end{equation}

\noindent
In the seminal \cite{gogioso_hyper} it was argued that some completion to a normalised process would indeed be possible in principle, but it was not known whether this be possible \emph{within} the theory of density hypercubes.
We now process to show that it is possible indeed, giving hyper-decoherence the operational interpretation of a probabilistic process.

\begin{prop}\label{completehypdec}
    The hyper-decoherence map is sub-normalised within the theory of density hypercubes.
    For qubits, its completion to a normalised process is given as follows:
    \begin{equation}\label{completionhypdec}
        \tikzfigscale{1}{figs/completionhypdec}
    \end{equation}
    where the white dot is a spider in the Pauli Z basis and the black dot is a phased spider in the Pauli X basis.
    For higher dimensions, its completion to a normalised process is given as follows:
    \begin{equation}\label{completionhypdecgeneral}
        \tikzfigscale{1}{figs/completionhypdecgeneral3}
    \end{equation}
    where the white dot is a spider in the computational basis, $\mathcal{K}(\raisebox{-1pt}{\scalebox{1.5}{$\circ$}})$ is the set of computational basis states and the black dot is a spider in any Fourier basis.
\end{prop}

Proposition \ref{completehypdec} above shows that the theory of density hypercubes splits into two ``sectors'': a quantum sector accessed by hyper-decoherence and another sector referred to as \emph{the Beyond}. It also shows that hyper-decoherence occurs probabilistically as one outcome of the above normalised process.

Discarding the map which completes the hyper-decoherence map (the second part of \eqref{completionhypdecgeneral}) gives an important effect which give a special name and symbol.

\begin{defn}
    The \emph{Unspeakable Horror from Beyond} (UHfB) is the effect completing the quantum discarding map (LHS of \ref{hypdecsubcaus}) to the full discarding map of density hypercubes. In the qubit case we have:
    \begin{equation}\label{UHfB}
        \tikzfigscale{1}{figs/UHfBeffect}
    \end{equation}
    We adopt the same symbol for all dimensions.
\end{defn}

The importance of the UHfB is that it can be used to complete quantum measurements/POVMs to genuine measurements/POVMs on density hypercubes.
For example, the following completes a computational basis measurement on a qubit to a measurement of density hypercubes:

\begin{equation}\label{POVM}
    \tikzfigscale{1}{figs/POVM}
\end{equation}

\noindent
This completion is necessary for a meaningful operational perspective on the larger theory, but is not observable from within quantum theory.
Indeed, consider a generic quantum state, taking the following form \cite{gogioso_hyper}:

\begin{equation}
    \tikzfigscale{1}{figs/quantumstate}
\end{equation}

\noindent
This quantum state has probability zero of yielding the UHfB as a measurement outcome:

\begin{equation}
    \tikzfigscale{1}{figs/quantumnotbeyond}
\end{equation}

\noindent
The first equality is by Hopf rule and the second equality is due to the black $\pi$ dot being the scalar 0.

Although the computational basis states take the product form shown in \eqref{POVM}, this is not the case for generic quantum states. For example, the state on the left below is the quantum $\ket{+}$ state, while the state on the right is a post-quantum state (sent to the quantum $\ket{+}$ state by hyper-decoherence):

\begin{equation}\label{quantumplus}
    \tikzfigscale{1}{figs/quantumplus}
    \hspace{4cm}
    \tikzfigscale{1}{figs/plus}
\end{equation}

\section{Phase group}

In quantum theory, the \emph{phase group} for a decoherence map $\dec{\circ}: H \rightarrow H$ is the group formed by all invertible processes $U: H \rightarrow H$ which are quotiented away by $\dec{\circ}$, i.e. such that

\begin{equation}
    \dec{\circ} \cdot U = \dec{\circ}
\end{equation}

\noindent
This is formed by the \emph{phase gates}, the unitaries diagonal in the decoherence basis, and is isomorphic to a torus $T^{d-1}$.
For a qubit, it is the circle group $T^1$.

\begin{defn}
    The \emph{phase group} in density hypercubes for a decoherence map $\dec{\circ}$ is the group formed by all invertible processes $U: H \rightarrow H$ which are quotiented away by $\dec{\circ}$, i.e. such that:
    \begin{equation}
        \dec{\circ} \cdot U = \dec{\circ}
    \end{equation}
\end{defn}

\begin{prop}\label{proposition:doublephase}
    The processes obtained by doubling phase gates from quantum theory are always in the phase group:
    \begin{equation}\label{doublephase}
        \tikzfigscale{1}{figs/czdecomp2}
    \end{equation}
\end{prop}

\noindent
The maps in \eqref{doublephase} are natural choices, as they are sent to the usual phase gates of quantum theory by hyper-decoherence.
The phase gates of quantum theory themselves are not however in the phase group for density hypercubes, as they are not invertible within the larger theory.

Are there more elements in the phase group? Restricting temporarily to the case where $\dim{H}=2$, we can certainly find more:

\begin{equation}\label{cz}
    \tikzfigscale{1}{figs/czgate}
\end{equation}

\noindent
We have used the Euler decomposition of the Hadamard \cite{coecke_zx} on the right-hand side to demonstrate that this is indeed a valid map for density hypercubes.\footnote{Note that it is not possible to make the map from \eqref{cz} in the double dilation or double mixing, as the only caps available are those on $H$ and $H^*$: the conjugation would change the sign of the phases on one side, causing them to vanish.}
The map in \eqref{cz} is clearly invertible, and one can check with ease that it is erased by hyper-decoherence.
The map also resembles the controlled-Z gate on two qubits, suggesting that there be an entire additional family of maps just like it living in the phase group.

\begin{prop}\label{proposition:czphase}
    The following maps are in the phase group of density hypercubes for $\dim{H}=2$:
    \begin{equation}\label{czphase}
        \tikzfigscale{1}{figs/czphase}
    \end{equation}
\end{prop}

\noindent
It is notable that the maps \eqref{czphase} are really a composition of the following two maps:

\begin{equation}\label{phasegad}
    \tikzfigscale{1}{figs/czdecomp}
    \hspace{5cm}
    \tikzfigscale{1}{figs/czdecomp2}
\end{equation}

It is then the maps on the LHS of \eqref{phasegad} that introduce something new to the phase group, since the maps on the RHS are the doubled phase spiders from before.
These new maps have been appeared previously in the literature under the name of \emph{phase gadgets} \cite{kissinger_phasegadgets}.
It is easy to show that composing phase gadgets adds their phases:

\begin{equation}\label{phasegadadd}
    \tikzfigscale{1}{figs/phasegad}
\end{equation}

\noindent
Phase gadgets and doubled phase spiders commute, so that the phase group for density hypercubes in dimension $\dim{H} = 2$ is the torus $T^2 = S^1\cross S^1$.

\begin{prop}\label{proposition:generalphase}
    The following maps are in the phase group of density hypercubes for arbitrary dimensions, generalising \eqref{czphase}:
    \begin{equation}\label{generalphase}
        \tikzfigscale{1}{figs/generalbridgephase}
    \end{equation}
    Above, $\ket{\psi} = \sum_{k\in\mathcal{K}(\raisebox{-1pt}{\scalebox{1.1}{$\circ$}})}e^{i\theta_k}\ket{k}$ is a \raisebox{-1pt}{\scalebox{1.5}{$\circ$}}-phase state with $\theta_k=\theta_{k^{-1}}$ and $\littleantipode$ is the antipode.
\end{prop}

\section{Hyper-phase group}

Having characterised the phase group, we can now look at its post-quantum generalisation: the \emph{hyper-phase group}, defined below. Because decoherence maps are invariant under pre- or post-composition by hyper-decoherence, the hyper-phase group is always a subgroup of the phase group.

\begin{defn}
    The \emph{hyper-phase group} in density hypercubes for a hyper-decoherence map $\hypdec{\circ}$ it the group formed by all invertible processes $U: H \rightarrow H$ which are quotiented away by $\hypdec{\circ}$, i.e. such that
    \begin{equation}
        \hypdec{\circ} \cdot U = \hypdec{\circ}
    \end{equation}
\end{defn}

\begin{prop}\label{proposition:hypphasegroup2}
    The hyper-phase group of density hypercubes for $\dim{H}=2$ contains exactly the maps from \eqref{czphase}:
    \begin{equation}
        \tikzfigscale{1}{figs/czphase}
    \end{equation}
\end{prop}

\begin{prop}\label{proposition:hypphasegroupGeneral}
    The hyper-phase group of density hypercubes for arbitrary dimensions contains the maps from \eqref{generalphase}:
    \begin{equation}
        \tikzfigscale{1}{figs/generalbridgephase}
    \end{equation}
\end{prop}

\section{Double Dilation and Double Mixing}
\label{sec:doubledilationmixing}

In this section we look at how many of the post-quantum features from density hypercubes are also available in double dilation and double mixing.

To start with, a minor alteration of the proof given in \cite{gogioso_hyper} can be used to show that double dilation and double mixing are also probabilistic theories.

\begin{prop}
    The decoherence maps of density hypercubes are also decoherence maps for double dilation and for double mixing, i.e. they send double dilated and double mixed systems to classical systems.
\end{prop}

\noindent
However, the hyper-decoherence maps from density hypercubes are not maps in double dilation or double mixing.  The most likely candidate candidate would be the following map (which however does not exist for double mixing):

\begin{equation}\label{possiblehypdec}
    \tikzfigscale{1}{figs/possiblehypdec}
\end{equation}

\begin{prop}\label{nohypdec}
    Map \eqref{possiblehypdec} does not give a hyper-decoherence map for double dilation.
\end{prop}

\noindent
A less rigorous but more straightforward way of seeing that map \eqref{possiblehypdec} cannot possibly be a hyper-decoherence map is to note that it erases the doubled phased spiders.
In fact, it is easy to show that the hyper-phase group would be limited to doubled unitaries.

\begin{prop}\label{invertible}
    In double dilation and double mixing the invertible maps are all doubled unitaries.
\end{prop}

\noindent
Proposition \ref{invertible} above immediately implies that the phase group for double dilation and double mixing is exactly the same phase group of quantum theory.
In particular, double dilation is not the same theory as density hypercubes.
Furthermore, even if double dilation and/or double mixing \emph{did} possess hyper-decoherence maps, they would not quotient away any non-trivial phases: it may ultimately turn out that one or both are post-quantum theories, but uninteresting ones at best.

\section{Conclusion}

In this work, we have conclusively shown that density hypercubes possess hyper-decoherence maps with a well-defined operational interpretation.
We have studied the associated phase group and we have compared our results with analogous statements for double dilation and double mixing.

The probabilistic nature of hyper-decoherence in density hypercubes presents a concrete way around the no-go theorem of Lee and Selby \cite{lee_nogo}.
Simply dropping the constraint that the hyper-decoherence be deterministic allowed the formulation of an operational theory displaying genuinely post-quantum phenomena, together with a mechanism for quantum theory to arise as an effective sub-theory.

A number of questions remain open.
There is initial evidence suggesting that density hypercubes would exhibit higher-order interference \cite{gogioso_hyper}, but final confirmation will require a fully fledged simulation of the triple and quadruple slit interference experiments within density hypercubes.
This will be our first upcoming endeavour.
Confirming higher-order interference would then immediately lead to investigation of computational advantage in density hypercubes.

Gaining a better understanding of the structure of the Beyond would also be an interesting route forward. In general, it seems likely that the Beyond is a non-trivial sector of density hypercubes, which could define its own operational theory, perhaps equivalent to quantum theory again. Investigations of this nature might also help shed some light on whether there are more elements of the hyper-phase group than we discovered in this work.

From a foundational perspective, we are also interested in exploring variations on the current formulation of the theory, e.g. by describing it directly from the perspective of quantum observers.
Preliminary investigations suggest that this would result in a theory with deterministic hyper-decoherence maps, but where ``pure'' quantum states would become fundamentally mixed as states of density hypercubes.

Finally, we hope to study the implications of probabilistic hyper-decoherence maps from a theory-independent perspective, refining the existing no-go result and computational advantage investigations.

\subsubsection*{Acknowledgements} JH would like to thank Matthew Wilson for helpful discussions and suggestions. This work was supported by University College London and the Engineering and Physical Sciences Research Council [grant number EP/L015242/1].
SG would like to thank Carlo Maria Scandolo for helpful discussions and acknowledge support from Hashberg Ltd.

\bibliographystyle{eptcs}
\bibliography{bibliography}

\appendix

\section{Proofs}

\subsection{Proof of Proposition \ref{completehypdec}}

\begin{proof}

    The second map of \eqref{completionhypdec} can be written with two black $\pi/2$ phases on the bridge and is therefore a valid map of density hypercubes. Applying the discarding maps we get:

    \begin{equation}\label{hypdecdecomp}
        \tikzfigscale{1}{figs/hypdecdecomp}
    \end{equation}

    \noindent
    For higher dimensions $d \geq 3$, the completion is slightly more complicated.
    Let $G$ be a finite abelian group on $d$ elements and \raisebox{-1pt}{\scalebox{1.5}{$\circ$}} correspond to the group element basis in the group algebra $\mathbb{C}[G] \cong \mathbb{C}^d$.
    Let \raisebox{-1pt}{\scalebox{1.5}{$\bullet$}} correspond to the Fourier basis for $G$, spanned by the (normalised adjoints of the) characters for $G$.
    Let $\mathcal{K}(\raisebox{-1pt}{\scalebox{1.5}{$\circ$}}) = \{\littlestate{k} : k \in G\}$ be the set of classical states for \raisebox{-1pt}{\scalebox{1.5}{$\circ$}}.
    Now consider the following CP map:

    \begin{equation}\label{characters}
        \tikzfigscale{1}{figs/completionhypdecgeneral4}
    \end{equation}

    \noindent
    It is easy to check that this gives a completion of the hyper-decoherence map to a normalised CP map, but it is not immediately clear that this is a valid map in the theory of density hypercubes.
    Indeed, it is not clear that it respects the symmetry required for maps of density hypercubes.
    However, writing $\bar{k}$ for the inverse of $k\in G$ we note the following equality
    
    \begin{equation}
        \tikzfigscale{1}{figs/bridgegroupequality}
    \end{equation}

    \noindent
    Which in turn implies:
    \begin{equation}
        \tikzfigscale{1}{figs/symmetricsum}
    \end{equation}
    
    \noindent
    Thus we can write:
        \begin{equation}
        \tikzfigscale{1}{figs/symmetricsum2}
    \end{equation}

    \noindent
    The ``control state'' $C$ on the RHS above is formed as follows:

    \begin{equation}
        \tikzfigscale{1}{figs/control}
    \end{equation}

    \noindent
    where $V: \mathbb{C}^2 \morph{} \mathbb{C}[G]$ is the isometry in $\fdhilb$ defined by $\ket{0}\mapsto \ket{0_G}, \ket{1}\mapsto \ket{\lambda}$.

\end{proof}

\subsection{Proof of Proposition \ref{proposition:doublephase}}

\begin{proof}

    By spider fusion, it is clear that the doubled phase gates are erased by hyper-decoherence:

    \begin{equation}
        \tikzfigscale{1}{figs/phaseerase}
    \end{equation}

\end{proof}

\subsection{Proof of Proposition \ref{proposition:czphase}}

\begin{proof}

    Showing that the maps \eqref{czphase} are erased by decoherence is a simple application of the Hopf rule.
    The harder part is showing that the maps \eqref{czphase} exist in density hypercubes: this boils down to showing that they have a symmetric expansion about the bridge, just as we have previously shown for \eqref{cz}.
    This expansion can be found by taking the square root of the following map in $\fdhilb$:

    \begin{equation}
        M(\alpha) \ = \ \tikzfigscale{1}{figs/Mmatrix} \sqrt{2}e^{i\alpha/2}
        \begin{pmatrix}
        \cos{\alpha/2} & 0 \\
        0 & -i\sin{\alpha/2}
        \end{pmatrix}
    \end{equation}

    \noindent
    where the matrix on the RHS is written in the Pauli X basis.
    Since $M(\alpha)$ is diagonal in the Pauli X basis, a square root $R(\alpha)$ is guaranteed to exist, itself diagonal in the Pauli X basis and thus self-transpose in Pauli X basis.
    Therefore we can safely write the square root on either side of the bridge as follows:

    \begin{equation}\label{Mbridge}
        \tikzfigscale{1}{figs/Mbridge}
    \end{equation}

\end{proof}

\subsection{Proof of Proposition \ref{proposition:generalphase}}

\begin{proof}

    The doubled phase spiders contribute the usual quantum phase group $T^{d-1}$.
    We also have the following maps:

    \begin{equation}\label{generalbridgephase}
        \tikzfigscale{1}{figs/generalbridgephase}
    \end{equation}

    \noindent
    where $\psi$ is a \raisebox{-1pt}{\scalebox{1.5}{$\circ$}}-phase state and $\littleantipode$ is the antipode.
    One can check that these maps are erased by decoherence and thus are in the phase group.
    All that is left to check is that they are normalised, invertible and that they exist in the theory of density hypercubes.
    Existence comes down to showing that they have a symmetric expansion about a bridge, generalising what happened in the proof of Proposition \ref{proposition:czphase}.
    Consider the following map in $\fdhilb$:

    \begin{equation}\label{subpart}
        \tikzfigscale{1}{figs/bridgephasesubpart}
    \end{equation}

    \noindent
    where we have expanded the phase state as its sum over \raisebox{-1pt}{\scalebox{1.5}{$\circ$}}-classical states and $\psi_k$ are complex numbers on the unit circle.
    One can see that \eqref{subpart} acts on \raisebox{-1pt}{\scalebox{1.5}{$\circ$}}-classical states as $\ket{g} \mapsto \sum_{k\in\mathcal{K}(\circ)} \psi_k \ket{g^{-1}k}$ (up to normalisation).
    Furthermore, \eqref{subpart} acts on \raisebox{-1pt}{\scalebox{1.5}{$\bullet$}}-classical states $\ket{\chi} = \sum_{g\in\mathcal{K}(\circ)}\chi(g)\ket{g}$ as follows (up to normalisation):

    \begin{equation*}
        \ket{\chi} \mapsto \left( \sum_{k\in\mathcal{K}(\circ)} \chi(k) \psi_k \right) \overline{\ket{\chi}}
    \end{equation*}

    \noindent
    where $\overline{\ket{\chi}} = \sum_{g\in\mathcal{K}(\circ)}\chi(g)^* \ket{g}$.
    (In the above, $\chi\in G^\wedge$ where $G^\wedge$ is the group of multiplicative characters for $G$.)
    We are thus able to expand \eqref{subpart} as a matrix in the \raisebox{-1pt}{\scalebox{1.5}{$\bullet$}} basis (again up to appropriate normalisation):

    \begin{equation*}
        \tikzfigscale{1}{figs/bridgephasesubpart2} = \sum_{\chi\in\mathcal{K}(\bullet)} \sum_{k\in\mathcal{K}(\circ)} \chi(k) \psi_k \overline{\ket{\chi}}\bra{\chi}
    \end{equation*}

    \noindent
    We want this matrix to have a square root and for this square root to be self-transpose with respect to the \raisebox{-1pt}{\scalebox{1.5}{$\bullet$}} basis.
    Since most entries of the matrix are zero, checking the existence of a square root comes down to looking at the sub-matrices for the terms $\ket{\chi}\overline{\bra{\chi}}$ and $\overline{\ket{\chi}}\bra{\chi}$, which take the form:

    \begin{equation}\label{weirdmatrix}
        \begin{pmatrix}
        0 & a\\
        b & 0
        \end{pmatrix}
    \end{equation}

    \noindent
    for some $a$ and $b$, where we have considered the case $\textrm{ord}(\chi)>2$ in $G^\wedge$.
    The case of $\textrm{ord}(\chi) = 2$ is trivial since it contributes a single non-zero diagonal element to the matrix, which can clearly be square rooted.
    The matrix \eqref{weirdmatrix} always has four square roots, since $a,b\neq 0$, but unless $a=b$ none of them are self-transpose in $\bullet$. In order to have $a=b$ we need the following to hold for each $\chi$:

    \begin{equation*}
        \sum_{k\in\mathcal{K}(\circ)} \chi(k) (\psi_k - \psi_{\bar{k}}) = 0
    \end{equation*}

    \noindent
    Note that the above is the Fourier transform of $f(k) = \psi_k - \psi_{\bar{k}}$ in the finite abelian group $G$.
    By inverting the transform, one sees that $\psi_k = \psi_{\bar{k}}$. This is trivially satisfied for those $k\in G$ such that $\textrm{ord}(k)=2$.
    Finally we show that maps \eqref{generalbridgephase} are closed under composition:

    \begin{equation}\label{generalgroupop}
        \tikzfigscale{1}{figs/generalgroupop}
    \end{equation}

    \noindent
    Above we have have used the Frobenius product $\ket{\psi\odot\phi} = \sum_{k\in\mathcal{K}(\circ)} \psi_k \phi_k \ket{k}$. This also shows that the maps \eqref{generalbridgephase} are invertible.
    As observed in the proof of Proposition \ref{proposition:czphase} before, the maps are also normalised, by the Hopf law.

\end{proof}

\subsection{Proof of Proposition \ref{proposition:hypphasegroup2}}

\begin{proof}

    One can check that the maps (\ref{phasegad}) are erased by hyper-decoherence:

    \begin{equation}
        \tikzfigscale{1}{figs/phasegad_hyp}
    \end{equation}

    \noindent
    where the final step follows by the Hopf law.
    On the other hand, the doubled phase spiders are \textit{not} erased by hyper-decoherence.
\end{proof}

\subsection{Proof of Proposition \ref{proposition:hypphasegroupGeneral}}

\begin{proof}

    The proof for Proposition \ref{proposition:hypphasegroup2} straightforwardly generalises to higher dimensions.

\end{proof}

\subsection{Proof of Proposition \ref{nohypdec}}

\begin{proof}

    An arbitrary tripartite pure state on systems $A, B$ and $C$ in double dilation can be written as follows:

    \begin{equation}
        \ket{\psi_{ABC}} = \sum_{ijk} c_{ijk} \ket{e_i^A e_j^B e_k^C}
    \end{equation}

    \noindent
    for $c_{ijk}\in\mathbb{C}$, where $\{\ket{e_i^A}\}$ forms an orthonormal basis for $A$, and similar for $B$ and $C$. By partial trace, a general state on $A$ in double dilation can be written as follows:

    \begin{align}
        \rho & = \sum_{pqrs} \bra{e_r^C e_p^B}\ket{\psi_{ABC}} \overline{\bra{e_s^C e_p^B}\ket{\psi_{ABC}}} \bra{e_s^C e_q^B}\ket{\psi_{ABC}} \overline{\bra{e_r^C e_q^B}\ket{\psi_{ABC}}} \\
        & = \sum_{ijklpqrs} c_{ipr} \ket{e_i^A} c^*_{ips} \overline{\ket{e_j^A}} c_{kqs} \ket{e_k^A} c^*_{lqr} \overline{\ket{e_l^A}}
    \end{align}

    \noindent
    Without loss of generality, we can consider the map \eqref{possiblehypdec} with \raisebox{-1pt}{\scalebox{1.5}{$\circ$}} associated to the basis $\{\ket{e_i^A}\}$. The result of applying the map to $\rho$ is the following state (written up to Choi-Jamio\l{}kowski isomorphism for convenience):

    \begin{equation}
        \sum_{ijpqrs} c_{ipr}c^*_{jps}c_{jqs}c^*_{iqr} \ket{e^A_i}\bra{e^A_j}
    \end{equation}

    \noindent
    We see that each of the coefficients is invariant under conjugation:
    \begin{equation}
        \left(\sum_{pqrs}c_{ipr}c^*_{jps}c_{jqs}c^*_{iqr}\right)^* = \sum_{pqrs} c^*_{ipr}c_{jps}c^*_{jqs}c_{iqr} = \sum_{pqrs} c_{ipr}c^*_{jps} c_{jqs}c^*_{iqr}
    \end{equation}

    \noindent
    where we relabelled $p$ and $q$ in the final step. As a consequence, the coefficients are all necessarily real and we do not recover all quantum states.

\end{proof}

\subsection{Proof of Proposition \ref{invertible}}

\begin{proof}

    The maps of double dilation can be written in the following form:

    \begin{equation}\label{ddmaps-invertible}
        \tikzfigscale{1}{figs/ddmaps}
    \end{equation}

    \noindent
    In order for maps in the form above to be invertible, the discarding maps need to be trivial (because of purity in $\cpm{\fdhilb}$):

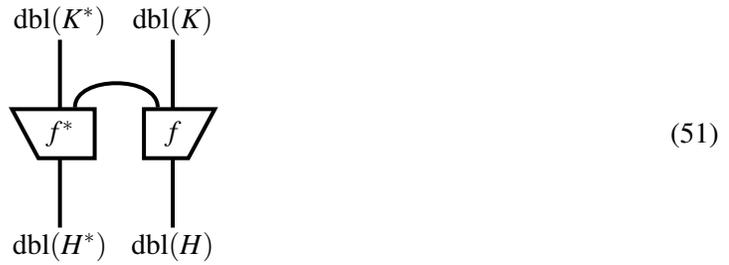
    \begin{equation}\label{ddmapspure-invertible}
        \tikzfigscale{1}{figs/ddmapspure}
    \end{equation}

    \noindent
    The diagram above is the doubled version of the following diagram in $\fdhilb$:

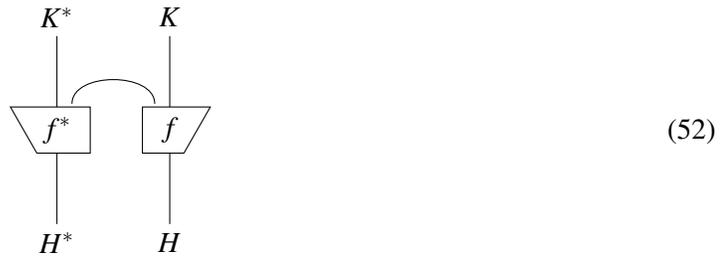
    \begin{equation}
        \tikzfigscale{1}{figs/ddmapspurefhilb}
    \end{equation}

    \noindent
    But the diagram above in $\fdhilb$ corresponds to \emph{another} CP map:

    \begin{equation}\label{cpmaps2}
        \tikzfigscale{1}{figs/cpmaps2}
    \end{equation}

    \noindent
    For the CP map \eqref{cpmaps2} above to be invertible, the discarding map must be trivial.
    This in turn implies that the bridge in \eqref{ddmapspure-invertible} must be trivial and hence that the original map \eqref{ddmaps-invertible} must take the following form if it is to be invertible:

    \begin{equation}
        \tikzfigscale{1}{figs/invertible}
    \end{equation}

    \noindent
    The map above is the double of a pure CP map and it is invertible exactly when $f$ is unitary.

\end{proof}

\end{document}

%% file: tikzstyles.tex
\tikzstyle{doubled}=[line width=1.5pt] 

\tikzstyle{dot}=[inner sep=0mm,minimum width=2mm,minimum height=2mm,draw,shape=circle]  
\tikzstyle{ddot}=[inner sep=0mm, doubled, minimum width=2.5mm,minimum height=2.5mm,draw,shape=circle]

\tikzstyle{pdot}=[inner sep=0mm, doubled, minimum width=2.5mm,minimum height=2.5mm,shape=circle]
\tikzstyle{phase dimensions}=[minimum size=6mm,font=\footnotesize,inner sep=0.2mm,outer sep=-2mm]

\tikzstyle{phase dot}=[pdot,phase dimensions]
\tikzstyle{wphase dot}=[dot, phase dimensions]

\tikzstyle{hadamard}=[fill=white,draw,inner sep=0.6mm,font=\footnotesize,minimum height=4mm,minimum width=4mm]

\tikzstyle{anti} = [fill=white,draw,inner sep=0.6mm,font=\footnotesize,minimum height=3mm,minimum width=3mm]

\tikzstyle{triang}=[regular polygon,regular polygon sides=3,draw,scale=0.75,inner sep=-0.75pt,minimum width=9mm,fill=white,regular polygon rotate=180]
\tikzstyle{triang_lesssep}=[regular polygon,regular polygon sides=3,draw,scale=0.75,inner sep=-4pt,minimum width=9mm,fill=white,regular polygon rotate=180, text depth=4mm]
\tikzstyle{triangdag}=[regular polygon,regular polygon sides=3,draw,scale=0.75,inner sep=-0.5pt,minimum width=9mm,fill=white]

\makeatletter
\newcommand{\boxshape}[3]{%
\pgfdeclareshape{#1}{
\inheritsavedanchors[from=rectangle] 
\inheritanchorborder[from=rectangle]
\inheritanchor[from=rectangle]{center}
\inheritanchor[from=rectangle]{north}
\inheritanchor[from=rectangle]{south}
\inheritanchor[from=rectangle]{west}
\inheritanchor[from=rectangle]{east}
\backgroundpath{
\southwest \pgf@xa=\pgf@x \pgf@ya=\pgf@y
\northeast \pgf@xb=\pgf@x \pgf@yb=\pgf@y

\@tempdima=#2
\@tempdimb=#3

\pgfpathmoveto{\pgfpoint{\pgf@xa - 5pt + \@tempdima}{\pgf@ya}}
\pgfpathlineto{\pgfpoint{\pgf@xa - 5pt - \@tempdima}{\pgf@yb}}
\pgfpathlineto{\pgfpoint{\pgf@xb + 5pt + \@tempdimb}{\pgf@yb}}
\pgfpathlineto{\pgfpoint{\pgf@xb + 5pt - \@tempdimb}{\pgf@ya}}
\pgfpathlineto{\pgfpoint{\pgf@xa - 5pt + \@tempdima}{\pgf@ya}}
\pgfpathclose
}
}}

\boxshape{NEbox}{0pt}{5pt}
\boxshape{SEbox}{0pt}{-5pt}
\boxshape{NWbox}{5pt}{0pt}
\boxshape{SWbox}{-5pt}{0pt}
\boxshape{EBox}{-3pt}{3pt}
\boxshape{WBox}{3pt}{-3pt}
\makeatother

\tikzstyle{map}=[draw,shape=SEbox,inner sep=2pt,minimum height=6mm,fill=white]
\tikzstyle{mapdag}=[draw,shape=NEbox,inner sep=2pt,minimum height=6mm,fill=white]
\tikzstyle{maptrans}=[draw,shape=NWbox,inner sep=2pt,minimum height=6mm,fill=white]
\tikzstyle{mapconj}=[draw,shape=SWbox,inner sep=2pt,minimum height=6mm,fill=white]

\tikzstyle{dmap}=[draw,doubled,shape=SEbox,inner sep=2pt,minimum height=6mm,fill=white]
\tikzstyle{dmapdag}=[draw,doubled,shape=NEbox,inner sep=2pt,minimum height=6mm,fill=white]
\tikzstyle{dmaptrans}=[draw,doubled,shape=NWbox,inner sep=2pt,minimum height=6mm,fill=white]
\tikzstyle{dmapconj}=[draw,doubled,shape=SWbox,inner sep=2pt,minimum height=6mm,fill=white]

\makeatletter
\pgfdeclareshape{cornerpoint}{
\inheritsavedanchors[from=rectangle] 
\inheritanchorborder[from=rectangle]
\inheritanchor[from=rectangle]{center}
\inheritanchor[from=rectangle]{north}
\inheritanchor[from=rectangle]{south}
\inheritanchor[from=rectangle]{west}
\inheritanchor[from=rectangle]{east}
\backgroundpath{
\southwest \pgf@xa=\pgf@x \pgf@ya=\pgf@y
\northeast \pgf@xb=\pgf@x \pgf@yb=\pgf@y

\pgfmathsetmacro{\pgf@shorten@left}{\pgfkeysvalueof{/tikz/shorten left}}
\pgfmathsetmacro{\pgf@shorten@right}{\pgfkeysvalueof{/tikz/shorten right}}

\pgfpathmoveto{\pgfpoint{0.5 * (\pgf@xa + \pgf@xb)}{\pgf@ya - 5pt}}
\pgfpathlineto{\pgfpoint{\pgf@xa - 8pt + \pgf@shorten@left}{\pgf@yb - 1.5 * \pgf@shorten@left}}
\pgfpathlineto{\pgfpoint{\pgf@xa - 8pt + \pgf@shorten@left}{\pgf@yb}}
\pgfpathlineto{\pgfpoint{\pgf@xb + 8pt - \pgf@shorten@right}{\pgf@yb}}
\pgfpathlineto{\pgfpoint{\pgf@xb + 8pt - \pgf@shorten@right}{\pgf@yb - 1.5 * \pgf@shorten@right}}
\pgfpathclose
}
}

\pgfdeclareshape{cornercopoint}{
\inheritsavedanchors[from=rectangle] 
\inheritanchorborder[from=rectangle]
\inheritanchor[from=rectangle]{center}
\inheritanchor[from=rectangle]{north}
\inheritanchor[from=rectangle]{south}
\inheritanchor[from=rectangle]{west}
\inheritanchor[from=rectangle]{east}
\backgroundpath{
\southwest \pgf@xa=\pgf@x \pgf@ya=\pgf@y
\northeast \pgf@xb=\pgf@x \pgf@yb=\pgf@y

\pgfmathsetmacro{\pgf@shorten@left}{\pgfkeysvalueof{/tikz/shorten left}}
\pgfmathsetmacro{\pgf@shorten@right}{\pgfkeysvalueof{/tikz/shorten right}}

\pgfpathmoveto{\pgfpoint{0.5 * (\pgf@xa + \pgf@xb)}{\pgf@yb + 5pt}}
\pgfpathlineto{\pgfpoint{\pgf@xa - 8pt + \pgf@shorten@left}{\pgf@ya + 1.5 * \pgf@shorten@left}}
\pgfpathlineto{\pgfpoint{\pgf@xa - 8pt + \pgf@shorten@left}{\pgf@ya}}
\pgfpathlineto{\pgfpoint{\pgf@xb + 8pt - \pgf@shorten@right}{\pgf@ya}}
\pgfpathlineto{\pgfpoint{\pgf@xb + 8pt - \pgf@shorten@right}{\pgf@ya + 1.5 * \pgf@shorten@right}}
\pgfpathclose
}
}

\makeatother

\pgfkeyssetvalue{/tikz/shorten left}{0pt}
\pgfkeyssetvalue{/tikz/shorten right}{0pt}

\tikzstyle{kpoint common}=[draw,fill=white,inner sep=1pt,minimum height=4mm]
\tikzstyle{kpoint}=[shape=cornerpoint,shorten left=5pt,kpoint common]
\tikzstyle{kpoint adjoint}=[shape=cornercopoint,shorten left=5pt,kpoint common]
\tikzstyle{kpoint conjugate}=[shape=cornerpoint,shorten right=5pt,kpoint common]
\tikzstyle{kpoint transpose}=[shape=cornercopoint,shorten right=5pt,kpoint common]

\tikzstyle{kpointdag}=[kpoint adjoint]
\tikzstyle{kpointadj}=[kpoint adjoint]
\tikzstyle{kpointconj}=[kpoint conjugate]
\tikzstyle{kpointtrans}=[kpoint transpose]

\tikzstyle{big kpoint}=[kpoint, minimum width=1.0 cm, minimum height=2mm, inner sep=4pt, text depth=1.5mm]

 \tikzstyle{upground}=[circuit ee IEC,thick,ground,rotate=90,scale=1.5]
 \tikzstyle{downground}=[circuit ee IEC,thick,ground,rotate=-90,scale=1.5]

%% file: styles.tikzstyles
\tikzstyle{discarding}=[fill=white, draw=black, shape=circle, style=upground]
\tikzstyle{smalldiscarding}=[fill=white, draw=black, style=upground, scale=0.5]
\tikzstyle{backdiscard}=[fill=white, draw=black, shape=circle, style=downground, scale=0.5]
\tikzstyle{smallbackdiscard}=[fill=white, draw=black, shape=circle, style=downground, scale=0.5]
\tikzstyle{state}=[fill=white, draw=black, style=triang, tikzit shape=rectangle]
\tikzstyle{kstate}=[fill=white, draw=black, style=kpoint, tikzit shape=rectangle]
\tikzstyle{kstateconj}=[fill=white, draw=black, style=kpoint conjugate, tikzit shape=rectangle]
\tikzstyle{kstateBIG}=[fill=white, draw=black, style=big kpoint, tikzit shape=rectangle]
\tikzstyle{effect}=[fill=white, draw=black, style=triangdag]
\tikzstyle{keffect}=[fill=white, draw=black, style=kpoint adjoint]
\tikzstyle{keffectconj}=[fill=white, draw=black, style=kpoint transpose]
\tikzstyle{morphdag}=[style=mapdag]
\tikzstyle{morph}=[style=map]
\tikzstyle{morphtrans}=[style=maptrans]
\tikzstyle{morphconj}=[style=mapconj]
\tikzstyle{CPMmorph}=[style=dmap]
\tikzstyle{CPMmorphconj}=[style=dmapconj]
\tikzstyle{CPMmorphdag}=[style=dmapdag]
\tikzstyle{CPMmorphtrans}=[style=dmaptrans]
\tikzstyle{CPMstate} = [fill=white, draw=black, style = triang, doubled]
\tikzstyle{CPMstateBIG} = [fill=white, draw=black, style = triang_lesssep, doubled]
\tikzstyle{CPMkstate}=[fill=white, draw=black, style=kpoint, tikzit shape=rectangle, doubled]
\tikzstyle{CPMkstateconj}=[fill=white, draw=black, style=kpoint conjugate, tikzit shape=rectangle, doubled]
\tikzstyle{CPMkstateBIG}=[fill=white, draw=black, style=big kpoint, tikzit shape=rectangle, doubled]
\tikzstyle{CPMkeffect}=[fill=white, draw=black, style=kpoint adjoint, doubled]
\tikzstyle{CPMkeffectconj}=[fill=white, draw=black, style=kpoint transpose, doubled]
\tikzstyle{UHfB}=[fill=white, draw=black, style=triangdag, doubled, inner sep=-2pt]
\tikzstyle{leak}=[style=tinypoint, regular polygon rotate = -90]
\tikzstyle{leakfill}=[style=tinypoint, regular polygon rotate = -90,fill=black]
\tikzstyle{Z} = [style=dot, fill=green]
\tikzstyle{X} = [style=dot, fill=red]
\tikzstyle{black_dot} = [style=dot, fill=black]
\tikzstyle{white_dot} = [style=dot, fill=white]
\tikzstyle{qblack_dot} = [style=ddot, fill=black]
\tikzstyle{qwhite_dot} = [style=ddot, fill=white]
\tikzstyle{whitephase} = [style=wphase dot, fill=white]
\tikzstyle{qredphase} = [style=phase dot, fill=red]
\tikzstyle{qgreenphase} = [style=phase dot, fill=green]
\tikzstyle{had} = [style=hadamard, doubled]
\tikzstyle{box} = [style=hadamard]
\tikzstyle{classhad} = [style=hadamard]
\tikzstyle{antipode} = [style=anti]

\tikzstyle{dottededge}=[-, dotted]
\tikzstyle{double edge}=[-, style=doubled, draw=black, tikzit draw={rgb,255: red,191; green,0; blue,64}]